\documentclass[aps,pra,twocolumn,superscriptaddress,floatfix]{revtex4-2}

\usepackage{hyperref}
\hypersetup{colorlinks=true,linkcolor=blue,urlcolor=blue,citecolor=blue}

\usepackage{bbm}

\usepackage{physics}
\usepackage{bm}	
\renewcommand{\vec}[1]{\bm{#1}}

\usepackage{placeins}
\usepackage{afterpage}
\usepackage{environ}
\usepackage{fullpage}
\usepackage{amssymb,amsmath}
\usepackage[utf8x]{inputenc}
\usepackage[T1]{fontenc}
\usepackage{siunitx}
\usepackage[version=3]{mhchem}
\usepackage[dvipsnames]{xcolor}
\usepackage{xfrac}
\newcommand\captionof[1]{\def\@captype{#1}\caption}

\usepackage{natbib}
\usepackage{graphicx}
\makeatletter
\def \figwidth{\columnwidth}
\makeatother

\usepackage{environ}
\makeatletter
\newwrite\remember@figures
\AtBeginDocument{\InputIfFileExists{\jobname.dft}{}{}\immediate\openout\remember@figures=\jobname.dft}
\AtEndDocument{\immediate\closeout\remember@figures}
\NewEnviron{dfigure}[1]{%
    \immediate\write\remember@figures{%
        \noexpand\rememberfigure{#1}{\unexpanded\expandafter{\BODY}}%
    }%
}
\NewEnviron{dfigure*}[1]{%
    \immediate\write\remember@figures{%
        \noexpand\rememberfiguretc{#1}{\unexpanded\expandafter{\BODY}}%
    }%
}
\newcommand{\placefigure}[2][tp]{%
    \csname remembered@figure@#2\endcsname{#1}%
}
\newcommand{\rememberfigure}[2]{%
    \global\@namedef{remembered@figure@#1}##1{%
        \begin{figure}[##1]#2\end{figure}%
    }%
}
\newcommand{\rememberfiguretc}[2]{%
    \global\@namedef{remembered@figure@#1}##1{%
        \begin{figure*}[##1]#2\end{figure*}%
    }%
}
\makeatother

\begin{document}
\title{Schrieffer-Wolff Transformations for Experiments: Dynamically Suppressing Virtual Doublon-Hole Excitations in a Fermi-Hubbard Simulator}
\newcommand{\harvard}{Department of Physics, Harvard University, 17 Oxford St., Cambridge, MA 02138, USA}
\newcommand{\tum}{Department of Physics and Institute for Advanced Study, Technical University of Munich, James-Franck-Str. 1, 85748 Garching, Germany}
\newcommand{\mcqst}{Munich Center for Quantum Science and Technology, Schellingstr. 4, 80799 M\"unchen, Germany}
\newcommand{\lmu}{Department of Physics and Arnold Sommerfeld Center for Theoretical Physics, Ludwig Maximilian University of Munich, Theresienstr. 37, 80333 M\"unchen, Germany}
\newcommand{\itamp}{ITAMP, Harvard-Smithsonian Center for Astrophysics, Cambridge, MA 02138, USA}
\newcommand{\princetonEE}{Department of Electrical Engineering, Princeton University, Princeton, New Jersey, 08544, USA}
\newcommand{\princetonCCM}{Princeton Center for Complex Materials, Princeton University, Princeton, New Jersey, 08540, USA}

\author{Anant Kale}
\affiliation{\harvard}
\author{Jakob Hendrik Huhn}
\affiliation{\lmu}
\author{Muqing Xu}
\author{Lev Haldar Kendrick}
\author{Martin Lebrat}
\affiliation{\harvard}
\author{Christie Chiu}
\affiliation{\harvard}
\affiliation{\princetonEE}
\affiliation{\princetonCCM}
\author{Geoffrey Ji}
\affiliation{\harvard}
\author{Fabian Grusdt}
\affiliation{\lmu}
\affiliation{\mcqst}
\author{Annabelle Bohrdt}
\affiliation{\harvard}
\affiliation{\itamp}
\author{Markus Greiner}
\email[Corresponding author: ]{greiner@physics.harvard.edu}
\affiliation{\harvard}


\begin{abstract}
In strongly interacting systems with a separation of energy scales, low-energy effective Hamiltonians help provide insights into the relevant physics at low temperatures. The emergent interactions in the effective model are mediated by virtual excitations of high-energy states: For example, virtual doublon-hole excitations in the Fermi-Hubbard model mediate antiferromagnetic spin-exchange interactions in the derived effective model, known as the $t-J-3s$ model. Formally this procedure is described by performing a unitary Schrieffer-Wolff basis transformation. In the context of quantum simulation, it can be advantageous to consider the effective model to interpret experimental results. However, virtual excitations such as doublon-hole pairs can obfuscate the measurement of physical observables. Here we show that quantum simulators allow one to access the effective model even more directly by performing measurements in a rotated basis. We
propose a protocol to perform a Schrieffer-Wolff transformation on Fermi-Hubbard low-energy eigenstates (or thermal states) to dynamically prepare approximate $t-J-3s$ model states using fermionic atoms in an optical lattice. Our protocol involves performing a linear ramp of the optical lattice depth, which is slow enough to eliminate the virtual doublon-hole fluctuations but fast enough to freeze out the dynamics in the effective model. We perform a numerical study using exact diagonalization and find an optimal ramp speed for which the state after the lattice ramp has maximal overlap with the $t-J-3s$ model state. We compare our numerics to experimental data from our Lithium-6 fermionic quantum gas microscope and show a proof-of-principle demonstration of this protocol. More generally, this protocol can be beneficial to studies of effective models by enabling the suppression of virtual excitations in a wide range of quantum simulation experiments.
\end{abstract}

\maketitle

\section{Introduction}
In recent years, quantum simulation experiments have been established as a valuable tool to investigate strongly correlated quantum many-body systems. 
Using the microscopic control of quantum simulators, Hamiltonians can be engineered in experiments and complex non-local correlators can be studied via site-resolved measurements.
Further, quantum simulators also enable basis transformations to be engineered to effectively perform measurements in different bases to reveal the underlying physics.
A canonical example of a basis transformation performed in cold-atom experiments is via time-of-flight imaging, which allows measurements in the momentum basis rather than the position basis.
In a similar vein, measurement of off-diagonal observables may be possible using a local unitary transformation realized via time-evolution under a quenched Hamiltonian such as in Refs.~\cite{Trotzky2008,Greif2013,Semeghini2021}. 
Measurements may also be performed in a randomized basis by applying local Haar-random unitary transformations to extract higher-order observables such as in Refs.~\cite{Brydges2019, Elben2018}.

In a quantum system with strong interactions, 
a basis transformation of particular interest is one that traces out the fast timescales in the system, revealing a low-energy effective model. 
Such effective models are frequently encountered in particle physics and studies of strongly correlated electronic systems, and can greatly facilitate physical insights, since they directly represent the relevant emerging interactions.

From a theoretical perspective, an effective model can be obtained from a system with a separation of energy-scales via the Schrieffer-Wolff transformation \cite{Schrieffer1966}, which involves a unitary transformation $\hat{U}=e^{i\hat{S}}$ to make the Hamiltonian block diagonal in the new basis. 
Importantly, new effective interactions emerge due to virtual excitations of high energy states when the original Hamiltonian is written in the dressed basis. 

In the context of quantum simulation, it can be advantageous to consider the effective model to interpret the experimental results. Examples include spin-exchange interactions in the strongly-interacting regime of the Bose- and Fermi-Hubbard model \cite{Duan2003}, as well as the
realization of an effective $U(1)$ gauge field, starting from a Bose-Hubbard model \cite{Yang2020}. However, if experimental measurements are performed in the original basis, the measured state could lie outside the low-energy sector due to quantum fluctuations (virtual excitations). Quantum simulation experiments are now exploring new regimes and access novel observables, such as spin-charge correlations \cite{Chiu2019Science,Koepsell2021}, with increasing accuracy. Details of the measurement procedure are therefore becoming more and more important. 
In order to accurately measure observables in the effective model, and thus avoid undesired virtual occupations, measurements should ideally be performed in the dressed basis, which is experimentally quite challenging in general.

In this work, we propose a protocol and demonstrate a proof-of-principle experiment to perform this required transformation from dressed to original basis, which dynamically eliminates virtual excitations thereby implementing approximately the Schrieffer-Wolff transformation ${e^{i\hat{S}}}$ to a quantum state before performing measurements. In particular, we focus on the case of using the Schrieffer-Wolff transformation to study the $t-J-3s$ model, which is the low-energy effective model of the doped Fermi-Hubbard Hamiltonian for large interaction strengths. However, our protocol can be readily generalized to other systems. 

The Fermi-Hubbard model, which is believed to constitute a minimal model for the physics of the cuprate materials, contains only two terms in the Hamiltonian – tunneling of fermions to neighboring lattice sites with amplitude $t$ and interaction energy $U$ between fermions on the same site:
\begin{equation} \label{eq:FermiHubbardHamiltonian}
    \hat{H}_{\rm{FH}} = -t \sum_{\langle i,j\rangle, \sigma = \uparrow, \downarrow}{\hat{c}^\dagger_{i,\sigma} \hat{c}_{j,\sigma} + h.c.} + U\sum_i{\hat{n}_{i,\uparrow} \hat{n}_{i,\downarrow} },
\end{equation}
where $\hat{c}_{i,\sigma}$ is a bare annihilation operator for a fermion with spin $\sigma$ in a Wannier orbital on lattice site $i$ and $\hat{n}_{i,\sigma} = \hat{c}^\dagger_{i,\sigma} \hat{c}_{i,\sigma}$ is the number operator. 
Despite its apparent simplicity, theoretical and numerical studies of the Fermi-Hubbard model have shown to be prohibitively difficult  in two or more dimensions due to its strong correlations and large entanglement. It can however be realized experimentally with ultracold atoms in optical lattices. Experimental studies with fermionic quantum gas microscopes have started to explore the physics of the Fermi-Hubbard model in regimes that are extremely difficult to simulate on classical computers \cite{Bohrdt2021Review, Mazurenko2017,Salomon2019, Koepsell2019,Chiu2018PRL,Ji2021, Nichols2019,Brown2019a}.

A low-energy effective Hamiltonian, called the $t-J-3s$ Hamiltonian, can be derived from the Hubbard Hamiltonian in the limit of large interaction energy $U \gg t$ via the Schrieffer-Wolff transformation \cite{MacDonald1988}. This allows one to exclude states with doubly occupied sites and thus significantly reduces the Hilbert space dimension of the model ($3^N$ vs. $4^N$ without taking symmetries into account). The derived $t-J-3s$ Hamiltonian is given by \cite{Auerbach1998}
\begin{equation} \label{eq:tJmodelHamiltonian}
    \hat{H}_{t-J-3s} = \hat{P}_s \Big( \hat{H}_t + \hat{H}_{\rm{QHM}} + \hat{H}_{3s} \Big) \hat{P}_s,
\end{equation}
where $\hat{P}_s$ is a projection operator onto the subspace containing no doubly occupied sites, and
\begin{align*}
    \hat{H}_t &= -t \sum_{\langle i,j\rangle, \sigma}{\tilde{c}^\dagger_{i,\sigma} \tilde{c}_{j,\sigma} + h.c.} \\
    \hat{H}_{\rm{QHM}} &= \frac{J}{2}\sum_{\langle i,j \rangle}{\Big(\hat{\vec{\tilde{S}}}_i \cdot \hat{\vec{\tilde{S}}}_j  - \frac{1}{4} \hat{\tilde{n}}_i \hat{\tilde{n}}_j \Big)} \\
    \hat{H}_{3s} &= -\frac{J}{8} \sum_{\langle i,j\rangle,\langle j, k\rangle}^{i\neq k}{\Big[\sum_\sigma{(\hat{\tilde{c}}_{i,\sigma}^\dagger \hat{\tilde{c}}_{k,\sigma} \hat{\tilde{n}}_j)} - \hat{\tilde{c}}_{i}^\dagger \vec{\sigma} \hat{\tilde{c}}_{k} \cdot \hat{\tilde{c}}_{j}^\dagger \vec{\sigma} \hat{\tilde{c}}_{j} \Big]}
\end{align*}
where $\hat{\tilde{c}}_{i,\sigma}$ is the dressed fermionic operator which is related to the bare fermionic operator via the Schrieffer-Wolff transformation, $\hat{\tilde{c}}_{i,\sigma} = e^{-i\hat{S}} \hat{c}_{i,\sigma} e^{i\hat{S}}$. The operator $\hat{S}$ is defined exactly as in \cite{MacDonald1988}. Here $J = 4 t^2/U$ is the super-exchange energy, $\hat{\vec{\tilde{S}}}_i = \hat{\tilde{c}}^\dagger_{i,a} \vec{\sigma}_{a,b} \hat{\tilde{c}}_{i,b}$ is a spin operator on site $i$ defined in terms of the dressed fermionic operators, and $\hat{\tilde{n}}_{i}$ is a dressed particle density operator on site $i$. 
Note that the three-site term $\hat{H}_{3s}$ is often neglected to simplify the effective model. However, we use the full $\hat{H}_{t-J-3s}$ Hamiltonian since $\hat{H}_{3s}$ appears at the same order as $\hat{H}_{\rm{QHM}}$ in $t/U$ in the expansion. 

The $t-J-3s$ model is also believed to contain much of the essential physics of the cuprate materials. In addition, the reduced Hilbert space dimension facilitates numerical studies of the model. There have been extensive theoretical and numerical studies of the $t-J$ and $t-J-3s$ model over the years, see e.g. \cite{Zhang1988,Dagotto1994,White1997,Corboz2014,Spalek2007,Wang2015a}, which renders quantum simulation of this model particularly interesting. 

Additionally, the $t-J-3s$ model does not display the quantum fluctuations in the Fermi-Hubbard model that cause virtual excitations of holes and doublons, i.e. doubly occupied sites. Close to half-filling, these fluctuations appear as virtual doublon-hole pairs that can still be observed experimentally, for example in density snapshots obtained by quantum gas microscopy \cite{Greif2016, Hartke2020}. The presence of virtual doublons and holes may obfuscate the density distribution of the system especially at low doping, such as in studies of single holes injected in a Mott insulator \cite{Ji2021, Vijayan2020,Koepsell2019} and studies of spin-charge as well as charge-charge correlations at finite doping \cite{Koepsell2021}. In such cases it can therefore be desirable to eliminate virtual density excitations by studying the simpler $t-J-3s$ model.

Here we propose a protocol to perform an approximate Schrieffer-Wolff transformation in a Fermi-Hubbard simulator via an optical lattice ramp which suppresses doublon-hole fluctuations and enables one to approximate correlators of the $t-J-3s$ model. The rest of this paper is organized as follows. In section \ref{sec:Protocol}, we explain the details of the protocol involving the optical lattice ramp. In section \ref{sec:DoublonElimination}, we present a simple two-site model which analytically explains the suppression of virtual doublon-hole pairs after the ramp. We also present numerical results and experimental data for larger system sizes. In section \ref{sec:SpinCorr}, we examine the spin correlations after the lattice ramp and its implications to thermometry. In section \ref{sec:LatticeRampUnitary}, we examine how the lattice ramp time-evolution operator effectively implements the approximate Schrieffer-Wolff transformation. In section \ref{sec:Conclusion}, we end with a conclusion and outlook.

\placefigure[!t]{figure1}


\section{Protocol}\label{sec:Protocol}
In the following section we explain the details of the protocol. 
Our goal is to dynamically map the dressed basis onto the original basis, so that we can measure observables of the effective model in the natural basis of the full model as long as their ground states are adiabatically connected. 
In the case of the $t-J-3s$ model, this corresponds to mapping $\hat{\tilde{c}}_{i,\sigma}$ onto  $\hat{c}_{i,\sigma}$ (see eq. \ref{eq:tJmodelHamiltonian}). Our protocol to perform this mapping ($\hat{\tilde{c}}_{i,\sigma} \rightarrow \hat{c}_{i,\sigma}$) involves a slow linear ramp of the optical lattice depth at the end of an experimental shot, followed by the usual fluorescence imaging sequence. The protocol is schematically illustrated in Fig. \ref{fig:ProtocolIntuition}. The linear ramp at a certain optimal speed acts as an approximate Schrieffer-Wolff transformation on low energy Fermi-Hubbard eigenstates and suppresses the doublon-hole fluctuations. This approximately maps the Fermi-Hubbard eigenstates onto corresponding eigenstates of the $t-J-3s$ model - as long as the ground states of the Hubbard and $t-J-3s$ models are adiabatically connected and there is no significant contribution from higher order terms in $t/U$ that were neglected in deriving the effective $t-J-3s$ model. These $t-J-3s$ model eigenstates can then be imaged in the natural basis of Fermi-Hubbard experiments. 

The lattice ramp results in a ramp of the Hamiltonian parameters. Increasing the strength of the lattice potential $V_0$ increases the localization of the Wannier functions, leading to a reduced tunneling amplitude $t$. Simultaneously, the increased localization also increases the interaction strength $U$. As a result, $U/t$ increases and $J/t$ decreases. In particular, the Hamiltonian parameters $t$ and $U$ scale as
\begin{align} 
\frac{t}{E_r} &\simeq  \frac{4}{\sqrt{\pi}} \left(\frac{V}{E_r}\right)^{3/4} \exp \left( -2 \sqrt{V/E_r} \right) \label{eq:TunellingVsLatticeDepth} \\
\frac{U}{E_r} &\simeq \sqrt{\frac{8}{\pi}} k_L a \left(\frac{V}{E_r}\right)^{3/4},  \label{eq:InteractionVsLatticeDepth}
\end{align}
where $V$ is the lattice depth, the recoil energy $E_r = \hbar^2 k_L^2/2m$ sets the energy scales of particles in the lattice and $k_L = 2\pi/\lambda_L$ is the laser wavevector. For a detailed derivation see Ref. ~\cite{Bloch2008, Tarruell2018} and references therein. 

\placefigure[!t]{figure2}

The effect of ramping the Hamiltonian parameters can be understood intuitively for the two extremal ramp speeds. Suppose the initial state is a low-energy Fermi-Hubbard eigenstate $\ket{\Psi}$. If $U/t$ is increased instantaneously to $U/t \rightarrow \infty$, the quantum state has no time to evolve and remains an eigenstate of the original Fermi-Hubbard model at finite $U/t$. This ``lattice freeze'' is generally performed in cold atom experiments prior to performing single-site resolved measurements, for example in \cite{Ji2021}.
In the opposite limit, if $U/t$ is increased fully adiabatically (i.e. slowly compared to all the many-body energy gaps), then the state adiabatically follows the instantaneous eigenstates towards the $U/t\rightarrow\infty$ limit which qualitatively changes the nature of the state in general (with the exception of the insulating half-filled ground state). 

For intermediate ramp speeds, the time-evolution of $\ket{\Psi}$ can be better understood by writing it in the basis of the $t-J-3s$ model as
\begin{equation} \label{eq:InitialStateIntJBasis}
\ket{\Psi} = \sum_i{\beta_i} \ket{\tilde{\phi}_i(t_0/U_0)},
\end{equation}
where $\{\ket{\tilde{\phi}_i(t_0/U_0)}\}$ are the Fock states of the dressed operators $\hat{\tilde{c}}_{j,\sigma}^{\dagger}$ and $t_0/U_0$ is the initial ratio of tunnelling to interaction strength. Note that $\ket{\tilde{\phi}_i(t/U)} = e^{-i\hat{S}(t/U)}\ket{\phi_i}$, where $\{\ket{\phi_i}\}$ are the Fock states of the original fermionic operators $\hat{c}_{j,\sigma}^{\dagger}$. Also note that $\ket{\tilde{\phi}_i(t/U \rightarrow 0)} = \ket{\phi_i}$ since $e^{-i\hat{S}(t/U)}\to \hat{\mathbbm{1}}$ as $t/U\to 0$. 
As the lattice depth is ramped up, the quantum state evolves in two ways: the coefficients $\{\beta_i\}$ evolve and the dressed basis states $\{\ket{\tilde{\phi}_i(t/U)}\}$ themselves evolve. 
If the lattice ramp is fast compared to $t$ and $J$, the coefficients of the dressed basis can be considered essentially frozen since their dynamics is governed by the $t-J-3s$ Hamiltonian. Additionally, if the ramp is slow with respect to $U$ which is the typical gap separating the low-energy sector from higher-energy sectors truncated out of the $t-J-3s$ Hamiltonian, the dressed basis states adiabatically flow towards $t/U \rightarrow 0$ and get mapped onto the original basis states $\{\ket{\phi_i}\}$. When both conditions on the ramp speed are satisfied, the final state can be approximated as:
\begin{equation} \label{eq:FinalStateAfterRamp}
    \ket{\Psi'} \approx e^{i\varphi_\Psi} \sum_i{\beta_i} \ket{\phi_i} = e^{i\varphi_\Psi} e^{i\hat{S}(t_0/U_0)}\ket{\Psi}
\end{equation}
where $e^{i\varphi_\Psi}$ describes an overall dynamical phase picked up during the ramp and $e^{i\hat{S}(t_0/U_0)}$ is the Schrieffer-Wolff transformation corresponding to the initial tunnelling to interaction ratio. 


Thus if the ramp speed is slow compared to $U$ but fast compared to $t$ and $J$, then the lattice ramp approximately acts as a Schrieffer-Wolff transformation on Fermi-Hubbard low-energy eigenstates, mapping them (up to order $t/U$) onto corresponding eigenstates of the $t-J-3s$ model written in the original basis. This allows for approximate measurements of observables of the  $t-J-3s$ model in the natural basis of the Fermi-Hubbard simulator.
\section{Eliminating Doublon-Hole pairs}\label{sec:DoublonElimination}
In this section, we examine how the lattice ramp suppresses virtual doublon-hole fluctuations. The population of virtual doublons tells us about how close the dressed operators $\tilde{c}_{i,\sigma}$ are to original operators $c_{i,\sigma}$. If the population of virtual doublons vanishes, the dressed basis gets mapped onto the original basis. We use the case of two fermions in a double-well potential (half-filling) to derive an approximate anayltical result for the doublon population after the lattice ramp. We also numerically study larger 1D and 2D systems and compare to experimental data for 2D systems.

For the double-well system, the Hamiltonian is illustrated in Fig. \ref{fig:Doublons}(a). The Hilbert space of interest with total spin $S = 0$ contains one spin up and one spin-down particle and is spanned by the four states $\{ \ket{\text{LL}}, \ket{\text{LR}}, \ket{\text{RL}}, \ket{\text{RR}} \}$ written following the convention of \cite{Murmann2015}. Further, examining the symmetries of the Hamiltonian (see Appendix \ref{app:LZ}), we can see that the dynamics of the system lie in a two dimensional Hilbert space spanned by the two spin-singlet states:
\begin{equation}
\begin{split}
    \ket{\Psi_{s}} &= \frac{\ket{\text{LR}} + \ket{\text{RL}}}{\sqrt{2}} ~\text{(singlet state)}\\
    \ket{\Psi_{dh}} &= \frac{\ket{\text{LL}} + \ket{\text{RR}}}{\sqrt{2}} ~\text{(symm. doublon-hole state)}
\end{split}
\end{equation}

For $U/t > 0$, the ground state of the double-well system is predominantly the singlet state $\ket{\Psi_s}$ with a small admixture of the symmetric doublon-hole state $\ket{\Psi_{dh}}$ of order $(t/U)^2$. If we start in the ground state with a finite $U/t$ and adiabatically increase $U/t \to \infty$, the system flows towards a purely singlet state with a vanishing doublon density. 

When the lattice depth is ramped up linearly in time (which is non-adiabatic, due to the exponential dependence of tunneling on lattice depth), the system does not perfectly follow the instantaneous ground state of the system. 
We can analytically solve the resulting dynamics in the two dimensional Hilbert space by making the following approximations. 
First, we Taylor-expand $U/t$ as a function of time to linear order and cast the problem into a Landau-Zener-like problem (see Appendix \ref{app:LZ}). Then we derive an analytical result for the state after a linear lattice ramp, when starting from the Fermi-Hubbard ground state with a given $U_0/t_0$. We compute the doublon density defined as $\langle \hat{\rho}_D\rangle = 1/N \sum_{i}^{N}{\langle n_{i\uparrow}n_{i\downarrow} \rangle}$ in the state after the ramp for a given ramp speed $\alpha \equiv d V /d\tau$:
\begin{equation} \label{eq:LZformula}
\langle \hat{\rho}_D\rangle \approx \frac{|\beta_{dh}^{(0)}|^2}{2} \frac{\alpha^2}{\alpha^2 + (\alpha^*)^2}.
\end{equation}
where $|\beta_{dh}^{(0)}|^2$ is the probability of the symmetric doublon-hole state for the initial Hamiltonian parameters $U_0$, $t_0$ and 
\begin{equation} \label{eq:AlphaStar}
    \alpha^* = U_0 \sqrt{V_0 E_r} \left[1 + 8 \Big(\frac{t_0}{U_0}\Big)^2 \right]
\end{equation}
is a critical lattice ramp speed where the doublon density has an inflection point. The Landau-Zener calculation predicts that in the adiabatic limit, i.e. ramp speed $\alpha \rightarrow 0$, the doublon density $\langle \hat{\rho}_D\rangle$ vanishes quadratically with $\alpha$. In the opposite limit of an instantaneous lattice ramp, $\alpha \rightarrow \infty$ the doublon density converges to its value in the ground state $|\beta_{dh}^{(0)}|^2/2$. The crossover in the behavior from quadratic to asymptotic occurs at a critical ramp speed $\alpha = \alpha^* \sim U \sqrt{V_0 E_r}$. 

In Fig.~\ref{fig:Doublons}(a), we compare the analytical result (dashed lines) with a numerical simulation of the full lattice ramp in the double well system without any approximations (solid lines) for two different initial values of $U_0/t_0 = 8$ (purple), and $U_0/t_0 = 16$ (brown). Here and in the following, the initial lattice depth is $V_0 = 8 E_r$ to match the experimental values of \cite{Mazurenko2017}, and the final lattice depth is $V_\text{final} = 48 E_r$. Increasing the value of $V_\text{final}$ does not affect the numerical results. The critical ramp speed $\alpha^*$ is marked with dash-dotted vertical lines. 

The analytical results match asymptotically with the full numerics in both limits $\alpha \rightarrow \infty$ (the doublon density is frozen to its value before the ramp) and $\alpha \rightarrow 0$ (the doublon density adiabatically vanishes).

For a given value of $U_0/t_0$, the crossover between adiabatic and frozen regime occurs at a larger $\alpha$ in the numerical results compared to the analytical formula. This discrepancy can be attributed to the different effective ramps realized by the two models: $U/t(\tau)$ increases linearly with time in the analytical approximation, whereas $V(\tau)$ increases linearly in the numerical model, resulting in an exponential ramp of $U/t(\tau)$ according to Eqs.~\ref{eq:TunellingVsLatticeDepth} and \ref{eq:InteractionVsLatticeDepth}. We expect the analytical result to be a good approximation to the numerics because most of the dynamics occur very early in the ramp where we can linearize the Hamiltonian parameters.

Next, we extend our numerical simulations to systems of up to 12 sites by computing the Fermi-Hubbard ground state using exact diagonalization (ED) and by evolving it according to Hamiltonian (\ref{eq:FermiHubbardHamiltonian}) with time-dependent parameters (\ref{eq:TunellingVsLatticeDepth}) and (\ref{eq:InteractionVsLatticeDepth}). Here and henceforth $U_0/t_0 = 8$ \cite{Mazurenko2017} (see Appendix \ref{app:Numerics} for details). Fig.~\ref{fig:Doublons}(b) shows results obtained at half-filling with one-dimensional chains and periodic boundary conditions (orange line) as well as two-dimensional $4 \times 3$ clusters with open boundary conditions (green line).

The doublon density after the lattice ramp is plotted as a function of the normalized ramp speed $\alpha/\alpha^*$ with $\alpha^*$ defined in Eq.~\ref{eq:AlphaStar} to compare timescales with the double-well case. For large ramp speeds $\alpha$, the doublon density reaches its ground-state value, which partially depends on the lattice coordination number and is different in the double-well, 1D
and 2D cases. Remarkably, the crossover speed $\alpha/\alpha^* \sim 1$ below which doublon density vanishes is in good qualitative agreement with the double well case. This suggests that the time and energy scales determining doublon suppression are mostly determined by the initial Hamiltonian parameters and are relatively independent of system size. We also performed 
numerics away from half-filling with low hole doping and found very similar results.


We also compare our numerics to experimental data from our Lithium-6 Fermi-Hubbard quantum simulator \cite{Parsons2015,Mazurenko2017}, shown with gray markers in Fig.~\ref{fig:Doublons}(b). 
The experimental data is for a spin-balanced mixture at half-filling with a system size of $\sim$ 370 sites. 
The tunnelling is set to $t=0.90(2)$ kHz and interaction strength is tuned using a Feshbach resonance to give $U/t = 8.1(2)$.
The system is loaded into a lattice of depth $7.5(1) E_r$, where $E_r = 25.6$ kHz, and the lattice is then ramped up by a factor of $8$ at varying ramp speeds to freeze the tunnelling. The fastest ramp shown corresponds to a ramp duration of $50\mu s$, while the slowest one corresponds to $10 ms$.
In the experimental snapshots, doublons and holes both appear as empty lattice sites as a result of the parity projection in the imaging scheme \cite{Bakr2009,Parsons2015}. The doublon density is extracted from the density snapshots by assuming that doublons and holes are equally likely since the system is at half-filling. 

The experimental data points are consistent with the 2D numerics (green line). We find that the experimentally obtained doublon density reduces with the ramp speed $\alpha$ on a similar timescale as the 2D numerics. For the experimental data, the density converges to about 1\% which is consistent with the imaging fidelity of approximately 98\% during these experiments.


\section{Spin correlations} \label{sec:SpinCorr}

In this section, we examine the spin correlation functions in the state after the lattice ramp and compare to correlations in the $t-J-3s$ model. As spin correlations are strongly affected by the presence of virtual doublon-hole excitations, they act as a proxy for how well the lattice ramp implements the Schrieffer-Wolff transformation on the initial state.

\placefigure[!t]{figure3}

In Fig.~\ref{fig:SpinCorr}(a), we show the nearest-neighbor spin correlation $\bra{\Psi_\text{ramp}} \hat{S}^z_i \hat{S}^z_j \ket{\Psi_\text{ramp}}_C$ evaluated on the state obtained after the lattice ramp as a function of the ramp speed $\alpha$ and doping $\delta$. To facilitate comparison between doping levels, we normalize correlations with the absolute value $|\bra{\Psi_{t-J}} \hat{S}^z_i \hat{S}^z_j \ket{\Psi_{t-J}}_C|$ obtained in the $t-J-3s$ model ground state, and normalize the ramp speed $\alpha$ by the critical ramp speed $\alpha^*$ in the double-well case (Eq.~(\ref{eq:AlphaStar})). The nearest-neighbor spin correlations are negative, which confirms the presence of anti-ferromagnetic order expected close to half-filling due to a positive superexchange coupling $J$.

At non-zero doping $\delta > 0$, the magnitude of the negative correlations shows a maximum at ramp speeds close to the critical ramp speed, $\alpha \sim \alpha^*$. There, correlations are about $\gtrsim 85\%$ of the value in the ground state of the $t-J-3s$ model for the finite system size of $4 \times 3$ sites. The presence of a maximum can be explained as the result of two competing effects. On the one hand, decreasing the ramp speed from the instantaneous ramp limit $\alpha \to \infty$ decreases the density of doublon-hole pairs, as observed in Fig.~\ref{fig:Doublons}. This effect contributes to decreasing the local magnetization on neighboring sites and therefore to increasing the magnitude of the spin correlations as $\alpha$ is decreased. On the other hand, in the adiabatic regime $\alpha < \alpha^*$ the quantum state follows the instantaneous ground state of the Hamiltonian during the ramp. In the fully adiabatic limit $\alpha \rightarrow 0$, the final state is described by a Hamiltonian with $U/t \rightarrow \infty$, i.e. $J/t \rightarrow 0$ and no spin correlations are present due to the Nagaoka effect \cite{Nagaoka1966}. In this case, increasing $\alpha$ towards $\alpha^*$ increases the magnitude of the correlations.

In the intuitive picture of Sec.~\ref{sec:Protocol}, Eq.~(\ref{eq:InitialStateIntJBasis}), a fast but finite-speed ramp leads to a transformation of the dressed eigenstates $\ket{\tilde{\phi}_i(t/U)}$ into bare eigenstates $\ket{\phi_i}$ with reduced doublon-hole pairs while keeping the coefficients $\beta_i$ unchanged, and the spin correlations of the ramped state approach the $t-J-3s$ model spin correlations. If the ramp speed $\alpha$ is further reduced, the coefficients $\beta_i$ start to evolve under the time-dependent Hamiltonian until the quantun states adiabatically follows the Hamiltonian ground state.

The half-filled case $\delta = 0$ [blue line in Fig.~\ref{fig:SpinCorr}(a)] is a special case: There, the tunneling $H_t$ and ring-exchange term $H_{3s}$ drop out of the $t-J-3s$ Hamiltonian and (\ref{eq:tJmodelHamiltonian}) becomes a Heisenberg Hamiltonian $\hat{H}_\text{QHM}$. Its ground state is independent of the magnitude of $J$ and shows constant, non-zero nearest-neighbor correlations, even in the limit $J/t \rightarrow 0$. When approaching the adiabatic limit $\alpha \rightarrow 0$, doublon-hole pairs get increasingly suppressed and there is no competing Nagaoka effect. Thus nearest-neighbor spin correlations monotonically increase in magnitude with decreasing ramp speed $\alpha$ and converge to their expectation value in the Heisenberg model.

Numerical simulations so far considered initial states at zero temperature, i.e. the ground states of the Fermi-Hubbard model. We now examine the case of finite-temperature ensembles in the Fermi-Hubbard model with temperatures $T$ below the initial interaction energy $U_0$ at half-filling. We perform full ED in a smaller system of $3 \times 3$ sites, simulate the time-evolution for the lowest $\sim 150$ eigenstates (from each magnetization sector) and average over the thermal ensemble by assigning appropriate Boltzmann weights to each state (see Appendix \ref{app:Numerics} for details). As shown in Fig.~\ref{fig:SpinCorr}(b) for $T/t \in [0, 0.5]$, temperature leads to an expected decrease of the nearest-neighbor spin correlations after the lattice ramp. Furthermore, spin correlations monotonically increase in magnitude by up to $25\%$ for decreasing ramp speed $\alpha/\alpha^*$, similar to the $T = 0$, $\delta = 0$ case in Fig.~\ref{fig:SpinCorr}(a). 

In quantum gas microscope experiments, single-site resolved measurements are performed after ramping up the lattice potential used for quantum simulation to much larger depths in order to ensure loss-less fluorescence imaging. Our simulations indicate that finite ramp speeds can lead to an overestimation of nearest-neighbor spin correlations and an underestimation of doublon-hole densities in Fermi-Hubbard systems close to half-filling. Taking into account these effects is therefore crucial to accurately estimate temperature in such systems, which often relies on comparing spin-spin or density-density correlation observables with numerical data obtained for example through Numerical Cluster Linked Expansion (NLCE) or Quantum Monte Carlo (QMC) methods \cite{Mazurenko2017,Hartke2020}.

\section{Analysis of the lattice ramp unitary} \label{sec:LatticeRampUnitary}
In this section we take a more careful look at the unitary operator describing time-evolution during the lattice ramp and how it is related to the Schrieffer-Wolff transformation. For a given ramp speed $\alpha$, we define the time-evolution operator of the lattice ramp as $\hat{U}_\text{ramp}(\alpha) = T\big[\exp{(-i \int_0^\tau{ d\tau' \hat{H}_{\rm{FH}}(\tau')})}\big]$.

\placefigure[!t]{figure4}

To probe the fidelity of the unitary operator at zero temperature, $T=0$, we compute the state overlap defined as $F = |\bra{\Psi_0^{tJ}}\hat{U}_\text{ramp}(\alpha)\ket{\Psi_0}|^2$ where $\ket{\Psi_0^{tJ}}$ is the $t-J-3s$ ground state, $\ket{\Psi_0}$ is the Fermi-Hubbard ground state (initial state) and $\hat{U}_\text{ramp}(\alpha)$ is the lattice ramp time evolution operator. This fidelity is shown in Fig. \ref{fig:LatticeRampUnitary}(a) as a function of ramp speed $\alpha$ and doping $\delta$ (solid lines). Dashed lines indicate the squared overlap of the initial state with the $t-J-3s$ ground state, i.e. $|\langle\Psi_0^{tJ}|\Psi_0\rangle|^2$. For computing the overlap, the $t-J-3s$ model ground state is written in terms of the bare fermionic operators $\hat{c}_{i,\sigma}$ rather than the dressed operators $\hat{\tilde{c}}_{i,\sigma}$, i.e. the $t-J-3s$ ground state written in the Fermi-Hubbard basis since this is the target state at the end of the ramp.

The results are qualitatively similar to the nearest-neighbor spin correlations shown in Fig. \ref{fig:SpinCorr}(a). Very fast ramps $\alpha \gg \alpha^*$ have no effect on the state and thus the overlap is given by that of the Fermi-Hubbard ground state with the $t-J-3s$ model ground state. For non-zero dopings $\delta > 0$, the overlap vanishes in the adiabatic limit $\alpha \ll \alpha^*$. Remarkably, it reaches a maximum at $\alpha^\text{opt} \approx 0.4 \alpha^*$, within an order of magnitude of the critical ramp speed predicted for the double-well case. The peak value is comparatively large (above 80\% for the 12-site system  considered here). As discussed in Sec. \ref{sec:SpinCorr}, the case of half-filling is special since the $t-J-3s$ model ground state becomes independent of $|J/t|$ then. As a result, the squared overlap with the $t-J-3s$ ground state monotonically increases to near unity with decreasing ramp speed.
These results quantitatively show that the lattice ramp protocol dynamically maps the Fermi-Hubbard ground state onto the $t-J-3s$ model ground state close to the critical ramp speed $\alpha^*$, allowing for measurements of $t-J-3s$ observables in the Fermi-Hubbard basis.

We now numerically examine the relation between the Schrieffer-Wolff unitary transformation $e^{i\hat{S}}$ and the lattice ramp time-evolution operator $\hat{U}_\text{ramp}$. 
Starting from a low energy eigenstate of the Fermi-Hubbard model $\ket{\Psi_n}$, by definition, the state after the ramp is given by $\ket{\Psi'} = \hat{U}_\text{ramp}\ket{\Psi_n}$. 
The $t-J-3s$ model eigenstate corresponding to $\ket{\Psi_n}$ is $e^{i\hat{S}} \ket{\Psi_n}$ up to order $(t/U)^2$ (see Sec.~\ref{sec:Protocol}). 
Thus we need $\hat{U}_\text{ramp} \ket{\Psi_n} \approx e^{i \varphi} e^{i\hat{S}} \ket{\Psi_n}$ up to a phase $\varphi$ for mapping the low-energy eigenstates of the Fermi-Hubbard model to the $t-J-3s$ model eigenstates.

In Fig.~\ref{fig:LatticeRampUnitary}(b), we numerically verify that this approximation is satisfied for the optimal ramp speed $\alpha^\text{opt} = 0.4 \alpha^*$ obtained in Fig.~\ref{fig:LatticeRampUnitary}(a) and small system size (8 sites, one-hole dopant), where the unitary operators can be fully computed. The fidelity $F_1 = |\bra{\Psi_n} e^{-i\hat{S}} \hat{U}_\text{ramp}^\text{opt}\ket{\Psi_n}|^2$ is plotted as a function of the energy $E_n$ (blue circles) to show the overlap between $\hat{U}_\text{ramp}$ and $e^{i\hat{S}}$ when applied on Fermi-Hubbard eigenstates. We find that the fidelity $F_1$ is quite high (above 90\% for this system) for all eigenstates within one tunneling energy of the ground state, $E_n - E_0 \leq t$. The fidelity $F_2 = |\bra{\Psi_n^{tJ}} \hat{U}_\text{ramp}^\text{opt}\ket{\Psi_n}|^2$ between the ramped state $\hat{U}_\text{ramp}^\text{opt}\ket{\Psi_n}$ and the corresponding $t-J-3s$ eigenstate (Fig.~\ref{fig:LatticeRampUnitary}(b), orange triangles) is also above 80\% for these low-energy eigenstates. This tells us that the lattice ramp protocol works well not just for the ground state but also for low-energy eigenstates and thermal states.

We note that $F_2$ has a wide spread in values - some eigenstates are mapped with very high fidelity (over 95\%) while others are much lower ($\sim$ 80\%). To understand the variance in $F_2$, we also plot $F_3 = |\bra{\Psi_n^{tJ}} e^{i\hat{S}}\ket{\Psi_n}|^2$ (green crosses) which compares how close $t-J-3s$ model eigenstates are to Fermi-Hubbard eigenstates after the Schrieffer-Wolff transformation. Since the Schrieffer-Wolff transformation is perturbation theory to order $(t/U)$ on an operator level, $F_3$ is expected to deviate away from unity for any finite $t/U$. Indeed we find that $F_3$ starts off high and shows a decreasing trend with energy with a significant variance. From the plot, the variance in $F_2$ can be explained by the variance in $F_3$ with a slight decrease in value from coming from the imperfect unitary $\hat{U}_\text{ramp}^\text{opt}$. The numerical evidence verifies that we can map low energy Fermi-Hubbard eigenstates (or thermal states) onto $t-J-3s$ model eigenstates (or thermal states) with high fidelity using the optimal lattice ramp.

\section{Conclusion and Outlook} \label{sec:Conclusion}
In this work, we proposed a protocol for measuring observables of low-energy effective models in quantum simulators by performing a ramp of Hamiltonian parameters that executes an approximate Schrieffer-Wolff basis rotation. We focused on the case of the $t-J-3s$ model derived from the Fermi-Hubbard model. In this case, the ramp of Hamiltonian parameters is performed by linearly increasing the lattice depth in time at an optimal ramp speed $\alpha^\text{opt}$. Using a simplified analytical model, numerical evidence as well as existing experimental data we demonstrated how the lattice ramp eliminates virtual doublon-hole fluctuations, increases spin-spin correlations in the system and executes an approximate Schrieffer-Wolff transformation, mapping the initial Fermi-Hubbard eigenstate onto the corresponding $t-J-3s$ model eigenstate. This mapping is possible in regimes where the (ground) states of the effective ($t-J-3s$) and microscopic (Hubbard) Hamiltonians are adiabatically connected.

While we discussed our protocol for studying observables in the effective model at equilibrium, we believe it can also be generalized for experiments studying non-equilibrium physics. For an out-of-equilibrium initial state, for the case of the Fermi-Hubbard and $t-J-3s$ models, we may be able to modify this protocol to separate the freezing of atomic motion from the elimination of doublon-hole fluctuations. For example, one could first use an energy offset on neighboring lattice sites to freeze the atomic motion as shown in \cite{Spar2021}, followed by the slow lattice ramp to eliminate doublon-hole fluctuations. 
The generalization to complex non-equilibrium initial states remains to be explored in future work.

The proposed protocol can also be applied in a larger range of systems whenever effective interactions are induced through virtual higher-order processes. Examples include U(1) lattice gauge theories in Bose-Hubbard systems \cite{Yang2020}, ring-exchange interactions \cite{Paredes2008}, and $\mathbbm{Z}_2$ lattice gauge theories with superconducting qubits \cite{Homeier2021PRB}.


\begin{dfigure}{figure1}
    \centering
    \noindent
    \includegraphics[width=\figwidth]{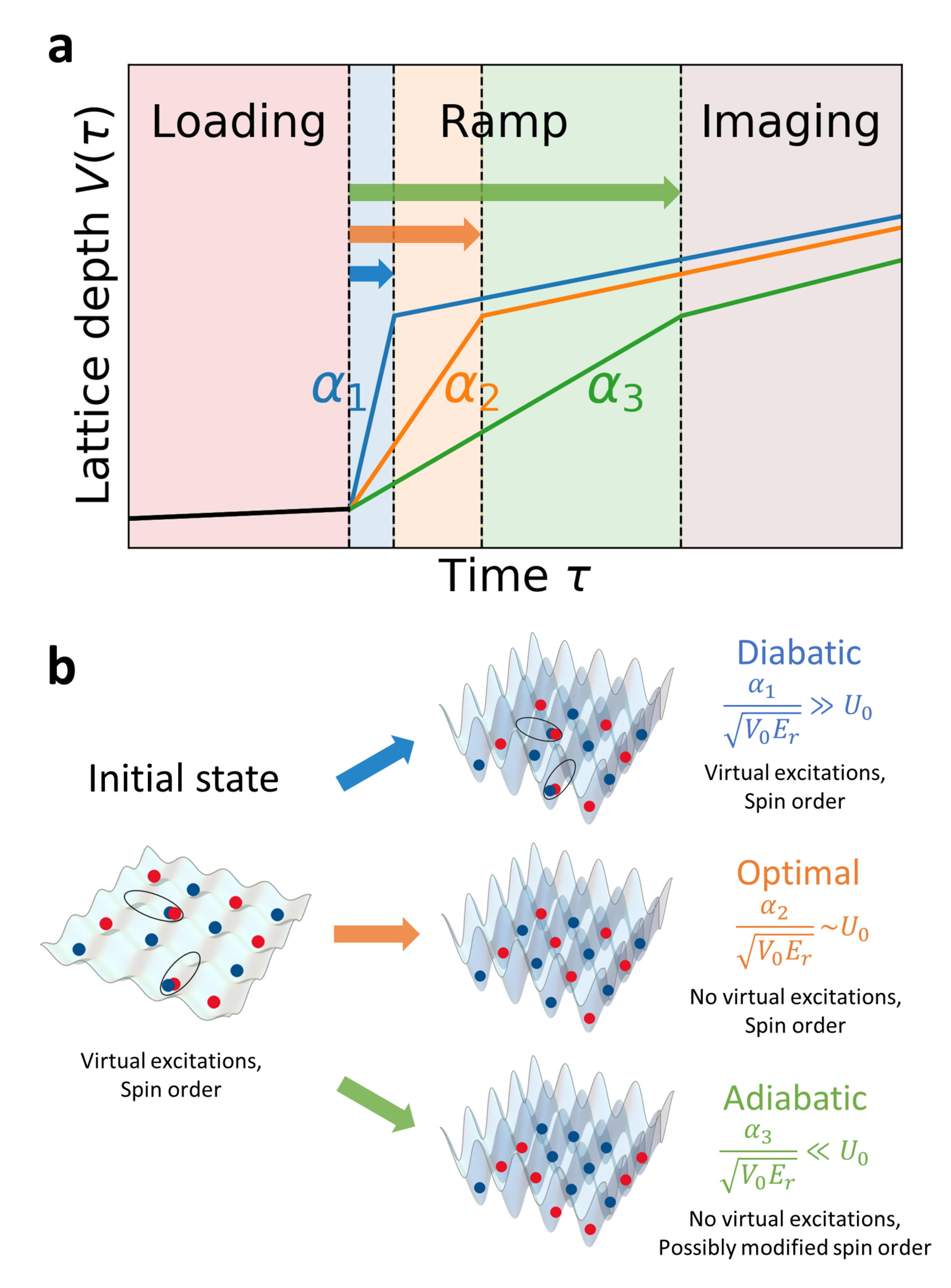}
    \caption{Schematic of the protocol: (a) The protocol involves a linear ramp of the optical lattice depth $V(\tau)$ at a rate $\alpha \equiv dV/d\tau$, followed by an imaging sequence to measure correlators. Colors indicate different ramp speeds.
    (b) If the lattice ramp is very fast, i.e. diabatic (blue), then the initial low-temperature Fermi-Hubbard state with doublon-hole virtual excitations and antiferromagnetic spin order is effectively frozen and remains the same after the ramp. 
    If the lattice ramp is very slow, i.e. adiabatic (green), the quantum state flows towards vanishing virtual excitations but also possibly modified spin order.
    In the intermediate regime (orange), for an optimal ramp speed virtual excitations are suppressed while also maintaining spin order, approximately mapping the initial state onto the effective model. See text for details. }
    \label{fig:ProtocolIntuition}
\end{dfigure}

\begin{dfigure*}{figure2}
    \centering
    \noindent
    \includegraphics[width=\linewidth]{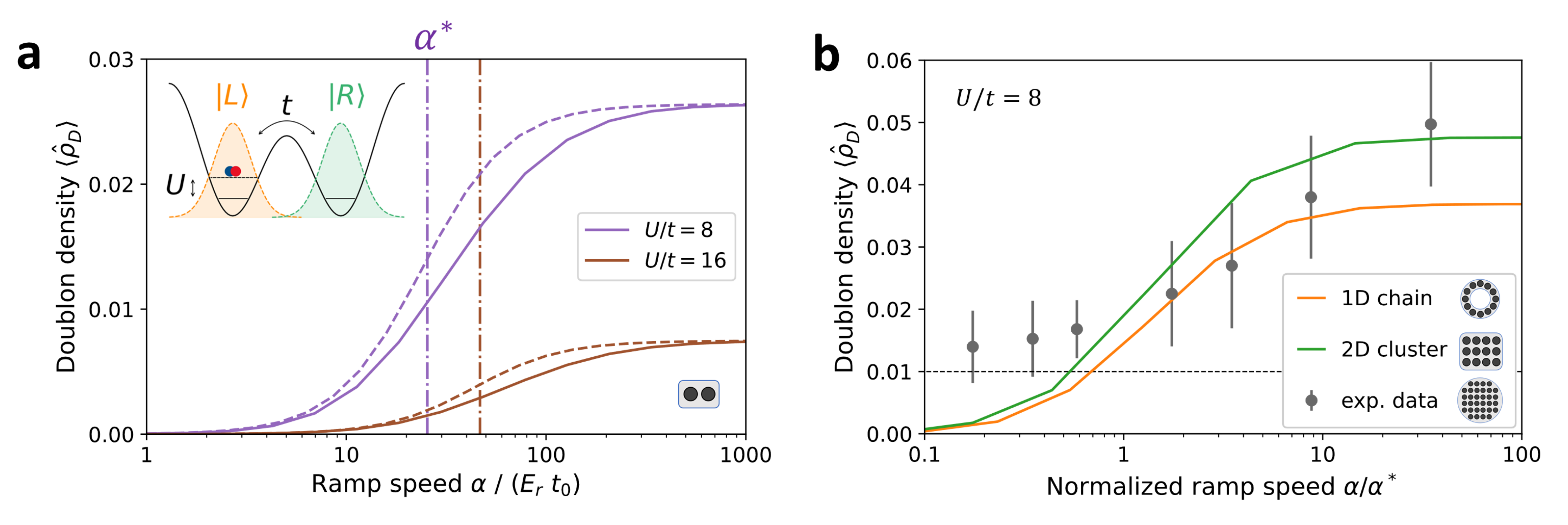}
    \caption{(a) Doublon density $\langle \hat{\rho}_D \rangle$ after the lattice ramp for the double-well case plotted as a function of ramp speed $\alpha$. Comparison of the analytical approximation of Eq.~\ref{eq:LZformula} (dashed lines) with numerical simulations (solid lines) for initial $U/t = 8$ (purple) and $U/t = 16$ (brown). Vertical dash-dotted lines indicate the critical ramp speed $\alpha^*$ from Eq.~\ref{eq:AlphaStar} below which the doublon density is strongly suppressed. The double-well Fermi-Hubbard Hamiltonian is schematically shown in the inset. (b) Doublon density $\langle \rho_D \rangle$ vs. normalized ramp speed $\alpha/\alpha^*$ for $U/t=8$. Solid lines are numerical results for a 12-site system at half-filling - 1D chain with periodic boundary conditions (orange line) and 2D cluster, $4 \times 3$ (green line). Experimental data points at half-filling shown in gray markers. The dashed line indicates the imaging fidelity limit to measuring doublon density in the experimental snapshots.}
    \label{fig:Doublons}
\end{dfigure*}

\begin{dfigure}{figure3}
    \centering
    \noindent
    \includegraphics[width=\figwidth]{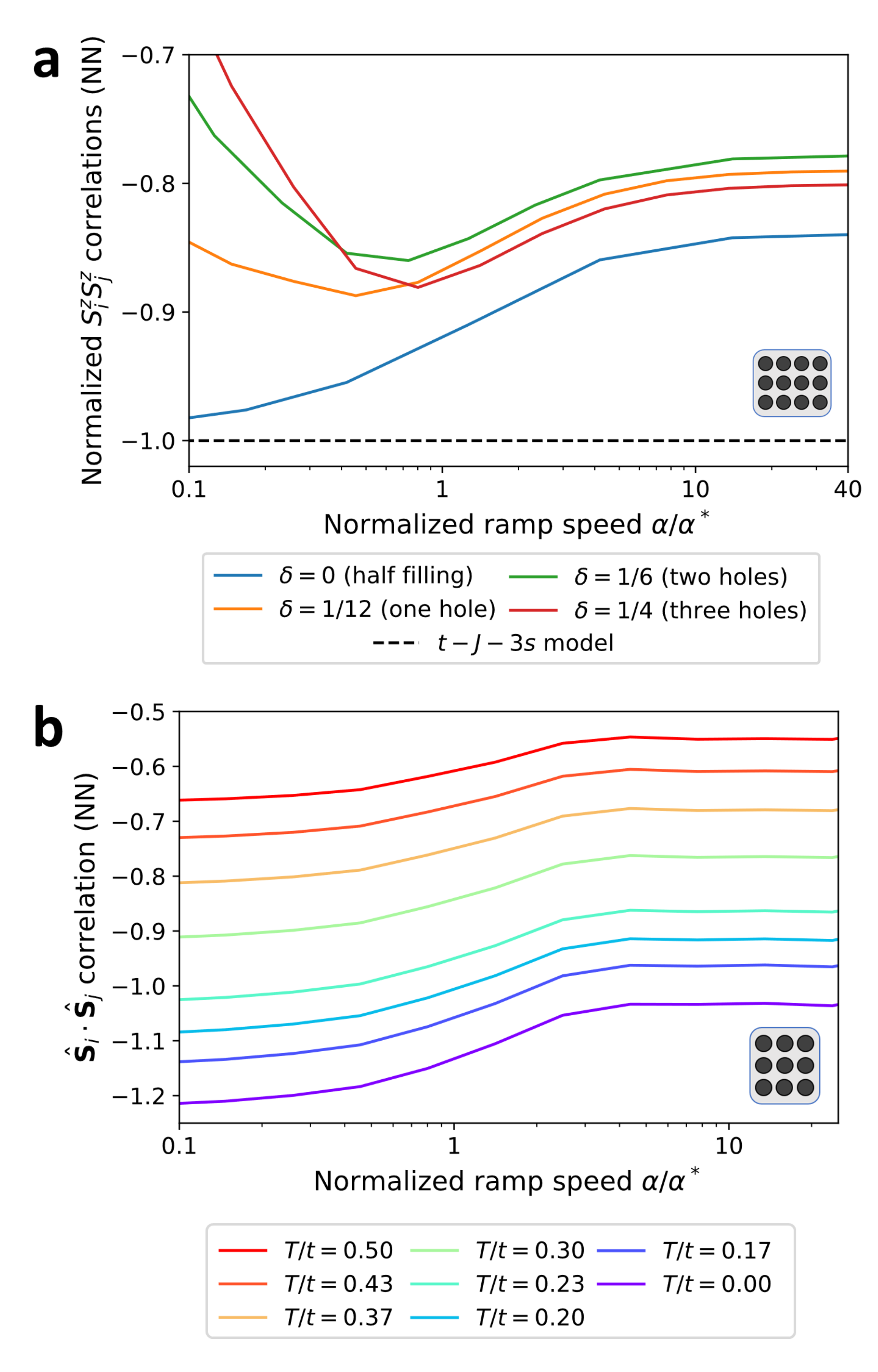}
    \caption{(a) Normalized spin correlation function ($z$ - component) for nearest neighbors 
$\langle \hat{S}^z_i \hat{S}^z_j \rangle_\text{ramp} / | \langle \hat{S}^z_i \hat{S}^z_j \rangle_{t-J}|$ 
plotted as a function of normalized ramp speed $\alpha/\alpha^*$ from numerical simulations at zero temperature. Colors indicate 
different doping levels. Dashed line indicates the normalized $t-J$ model correlator. Spin correlations are normalized by the magnitude of the $t-J-3s$ model correlators to plot different doping levels on the same plot. System size is 12-sites ($4\times3$).
    (b) Spin correlation function for nearest neighbors $\langle \hat{\vec{S}}_i \cdot \hat{\vec{S}}_j \rangle_C$ at finite temperature. Solid lines indicate numerical results for a 9-site ($3\times3$) system at half-filling with colors indicating different temperatures $T/t\in[0,0.5]$. 
$U_0/t_0 = 8$ for both plots.}
    \label{fig:SpinCorr}
\end{dfigure}

\begin{dfigure}{figure4}
    \centering
    \noindent
    \includegraphics[width=\figwidth]{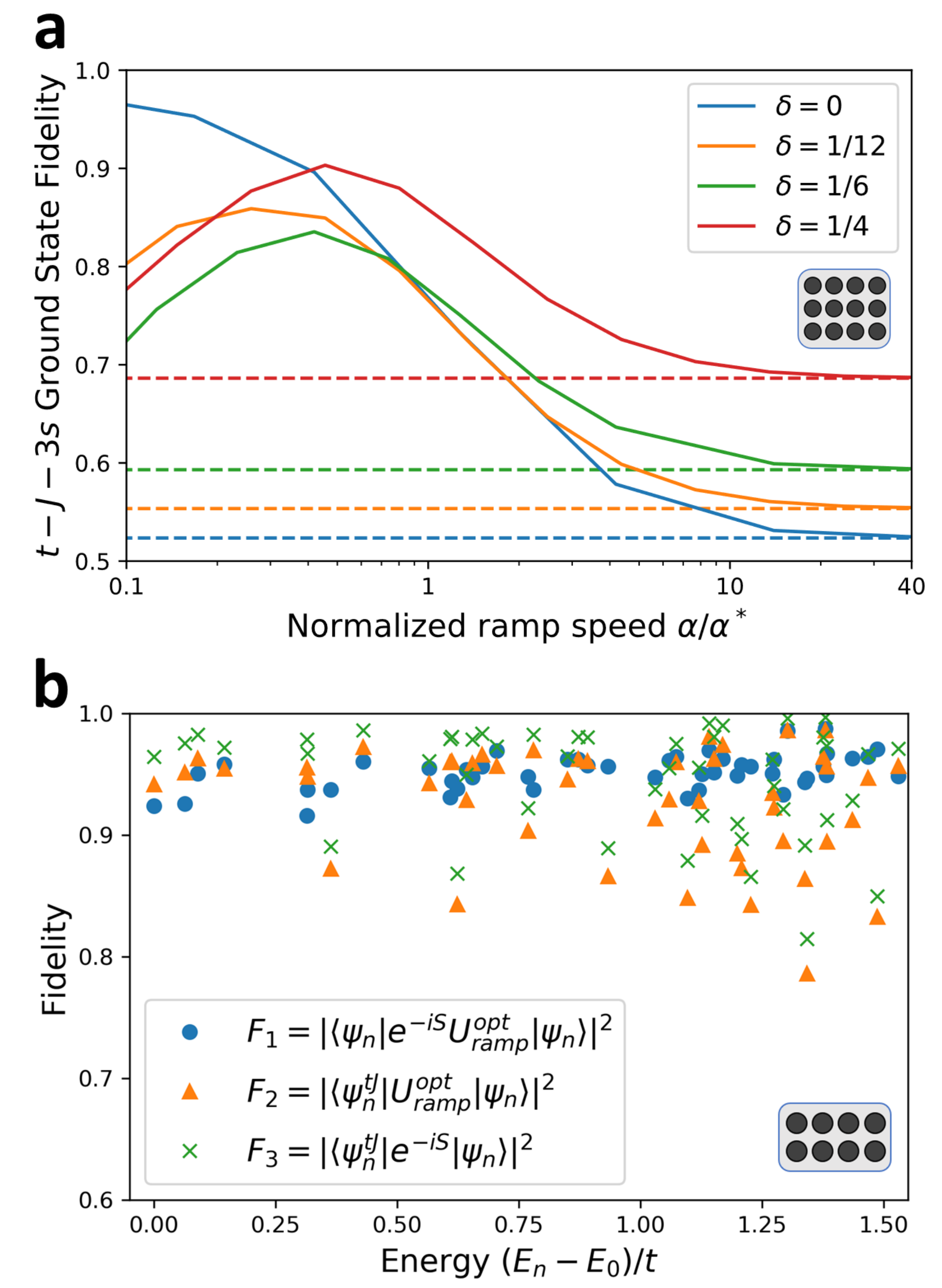}
    \caption{(a) Fidelity of mapping to the $t-J-3s$ model ground state starting from the Fermi-Hubbard ground state, defined as $F = |\bra{\Psi_0^{tJ}}\hat{U}_\text{ramp}(\alpha)\ket{\Psi_0}|^2$. Solid lines show numerical results for a 12 site cluster ($4 \times 3$) with colors indicating different doping levels. Dashed lines indicate the squared overlap of the initial state with the $t-J$ ground state, i.e. $|\langle \Psi_0^{tJ} | \Psi_0 \rangle|^2$. 
    (b) Comparison of the optimal lattice ramp unitary $\hat{U}_\text{ramp}^\text{opt}$ and the Schrieffer-Wolff unitary transformation $e^{i\hat{S}}$ for low energy eigenstates of the Fermi-Hubbard model. Fidelity $F_1$ (blue markers) shows how close $\hat{U}_\text{ramp}$ and $e^{i\hat{S}}$ are when acting on Fermi-Hubbard eigenstates, $F_2$ (orange triangles) shows the fidelity of mapping to a $t-J-3s$ eigenstate when starting from the corresponding Fermi-Hubbard low-energy eigenstate, and $F_3$ (green crosses) shows the squared overlap of $t-J-3s$ eigenstates and corresponding Fermi-Hubbard eigenstates. $U_0/t_0 = 8$ for both plots.}
    \label{fig:LatticeRampUnitary}
\end{dfigure}

\section{Acknowledgements}
We acknowledge funding from 
NSF grant nos. PHY-1734011, OAC-1934598 and OAC-1934714; 
ONR grants no. W911NF-11-1-0400 and N00014-18-1-2863; 
ARO grant no. W911NF-20-1-0163; 
ARO/AFOSR/ONR DURIP grant no. W911NF2010104; 
QSA Lawrence Berkeley Lab award no. DE-AC02-05CH11231;
the Gordon and Betty Moore Foundation grant no. 6791; 
the Harvard Quantum Initiative (HQI) Grauduate Fellowship (A.K.),
the NSF GRFP (L.H.K., C.C.), the DoD NDSEG (G.J.), 
and the NSF through a grant for the Institute for Theoretical Atomic, Molecular, and Optical Physics at Harvard University and the Smithsonian Astrophysical Observatory (A.B.);
the Swiss National Science Foundation and the Max Planck/Harvard Research Center for Quantum Optics (M.L.).
F.G. acknowledges funding by the Deutsche Forschungsgemeinschaft (DFG, German Research Foundation) under Germany's Excellence Strategy -- EXC-2111 -- 390814868, and from the European Research Council (ERC) under the European Union’s Horizon 2020 research and innovation programm (Grant Agreement no 948141) — ERC Starting Grant SimUcQuam. \\
Competing interests: M.G. is a co-founder and shareholder of QuEra Computing.

%

\clearpage
\appendix
\onecolumngrid
\section{Analytical Half-Landau Zener Formula} \label{app:LZ}
Our goal in this section is to analytically calculate the effect of a linear ramp of lattice depth on the quantum state of a two-site Fermi-Hubbard system. Starting from the ground state of the two-site system, as the lattice depth is increased, the Hamiltonian parameters $U$ and $t$ evolve in time and the state undergoes dynamics. If the change in Hamiltonian parameters is Taylor-expanded and linearized in time and the initial lattice depth is sufficiently deep, we can make use of the Landau-Zener formalism to solve the dynamics. 

However, our case is different from the usual Landau-Zener problem involving an avoided crossing because in our case, the Hamiltonian parameters always remain on one side of the avoided crossing ($U/t$ goes from $U_0/t_0>0$ to $U/t\to\infty$). The lattice ramp is thus more like the latter half of a Landau-Zener problem.

Consider the Hamiltonian of the two-site Fermi-Hubbard system at half-filling (2 particles):
\begin{equation}\label{eq:Hubbard2}
    \hat{H}=-t(\tau) \sum_{\sigma}\left(c_{L \sigma}^{\dagger} c_{R \sigma}^{} + c_{R \sigma}^{\dagger} c_{L \sigma}^{}\right)+U(\tau) \sum_{i=\{L,R\}} \hat{n}_{i \uparrow} \hat{n}_{i \downarrow}.
\end{equation}
We first find the ground state of the Hamiltonian for $U>0$. Considering the spin rotation and inversion symmetries of the system, the eigenstates of the Hamiltonian are also total spin eigenstates (singlet or triplet) and have a defined parity (even or odd). 
In the spin triplet sector, we find three eigenstates of the Hamiltonian, all with zero energy. In spin singlet sector, there are three states each with exactly one spin up particle and one spin-down particle. These states are spanned by the basis states:
\begin{equation}
    \begin{split}
        \mathcal{B}=\Big\{
        &\ket{\phi_0}=\frac{1}{\sqrt{2}}\left(\ket{LR} + \ket{RL}\right) 
        = \frac{1}{\sqrt{2}}\left(\hat{c}_{L\uparrow}^{\dagger}\hat{c}_{R\downarrow}^{\dagger} + \hat{c}_{R\uparrow}^{\dagger} \hat{c}_{L\downarrow}^{\dagger}\right)\ket{\text{vac}}
        \equiv \ket{\Psi_s},\\  
        &\ket{\phi_1}=\frac{1}{\sqrt{2}}\left(\ket{LL} + \ket{RR}\right)
        = \frac{1}{\sqrt{2}}\left(\hat{c}_{L\uparrow}^{\dagger}\hat{c}_{L\downarrow}^{\dagger} + \hat{c}_{R\uparrow}^{\dagger} \hat{c}_{R\downarrow}^{\dagger}\right)\ket{\text{vac}}
        \equiv \ket{\Psi_{dh}}, \\
        &\ket{\phi_2}=\frac{1}{\sqrt{2}}\left(\ket{LL} - \ket{RR}\right) 
        = \frac{1}{\sqrt{2}}\left(\hat{c}_{L\uparrow}^{\dagger}\hat{c}_{L\downarrow}^{\dagger} - \hat{c}_{R\uparrow}^{\dagger} \hat{c}_{R\downarrow}^{\dagger}\right)\ket{\text{vac}}
        \Big\}.
    \end{split}
\end{equation}
We have conveniently chosen these basis states to be parity eigenstates. We can see that $\ket{\phi_0}$ and $\ket{\phi_1}$ are even parity states while $\ket{\phi_2}$ is the only odd parity eignestate in this spin sector. Thus $\ket{\phi_2}$ must be an eigenstate of the Hamiltonian and has energy $E_2 = U$ (can be easily verified). The remaining two states are coupled by the Hamiltonian and form a simple two-level system. Diagonalizing the two-dimensional Hilbert space, we find the eigenstates of the Hamiltonian are given by:
\begin{align} \label{eq:TwoSiteEigenstates}
    \ket{\psi_0} &= \cos{\theta} \ket{\Psi_s} + \sin{\theta} \ket{\Psi_{dh}}, \qquad
    \qquad E_0 = \frac{1}{2}(U - \sqrt{U^2 + 16 t^2}), \\
    \ket{\psi_1} &= \sin{\theta} \ket{\Psi_s} - \cos{\theta} \ket{\Psi_{dh}}, \qquad
    \qquad E_1 = \frac{1}{2}(U + \sqrt{U^2 + 16 t^2})\\
    & \qquad\qquad \tan{\theta} = \frac{(-U + \sqrt{U^2 + 16 t^2})}{4t}
\end{align}
For repulsive interactions $U>0$, $E_0 < 0 < E_2 < E_1$. Thus $\ket{\psi_0}$ has the lowest energy out of all 6 states in the full two-site Hamiltonian, i.e. $\ket{\psi_0}$ is the ground state.

Now we consider time dependent tunneling $t(\tau)$ and interactions $U(\tau)$ dependent on the lattice depth $V(\tau)$. The initial state of the system is the ground state $\ket{\psi_0}$. As $U$ and $t$ vary in time $\tau$, the Hamiltonian mixes the state in the two-dimensional Hilbert space of $\ket{\phi_0}$ and $\ket{\psi_1}$. Let us write the solution as
\begin{equation}
        \ket{\Psi(\tau)} = a_0(\tau)\ket{\phi_0} + a_1(\tau)\ket{\phi_1}, \qquad
        a_0(0) = a_0^0 = \cos{\theta}, \quad a_1(0) = a_1^0 = \sin{\theta},
\end{equation}
where $\theta$ is the same is in eq. (\ref{eq:TwoSiteEigenstates}).

To cast the Hamiltonian into the same form as the Landau-Zener Hamiltonian, we rescale eq. (\ref{eq:Hubbard2}) by $t(\tau)$. Further, we also rescale time by $t(\tau)$ to recover the usual Schrodinger equation.
\begin{equation}
    i\frac{1}{t(\tau)}\partial_\tau \ket{\psi(\tau)} =\frac{\hat{H}(\tau)}{t(\tau)}\ket{\psi(\tau)}
        \Rightarrow i\partial_{\tilde{\tau}} \ket{\psi(\tilde{\tau})} = \tilde{H}(\tilde{\tau})\ket{\psi(\tilde{\tau})} 
\end{equation}
where $\tilde{H}(\tilde{\tau})=\frac{\hat{H}}{t(\tau)}$ and we introduce $\tilde{\tau}(\tau)$ such that $\frac{1}{t(\tau)}\partial_\tau\equiv\partial_{\tilde{\tau}}$.
The tunneling parameter is time dependent, therefore the re-scaling is not simply linear. However, since the majority of the change occurs when $U/t$ is small at the start of the ramp, we can linearize the rescaling by Taylor-expanding around $\tau=0$. 
\begin{equation}\label{timerescale}
    \frac{\partial\tilde{\tau}}{\partial\tau}=t(\tau) \Rightarrow \tilde{\tau}(\tau)=\int_{0}^\tau d\tau' t(\tau')\simeq t_0\tau, \quad t_0=t(0)
\end{equation}

From the Schrodinger equation, we get the following coupled differential equations:
\begin{equation} \label{eq:TwoSiteEOM}
    \begin{split}
        i\partial_{\tilde{\tau}}a_0 &= -g a_1\\
        i\partial_{\tilde{\tau}}a_1 &= - g a_0 + \frac{U\left(\tilde{\tau}\right)}{t\left(\tilde{\tau}\right)}a_1.
    \end{split}
\end{equation}
where $g = 2$ is the time-independent coupling parameter obtained by rescaling the Hamiltonian by the tunnelling strength. When $\hbar$ is set to $1$, $g$ and $2$ are interchangeable. However, it is useful to use $g$ instead of the numerical value $2$ in our solution because we can later restore $\hbar$ via $g = 2 \hbar$ and recover unit-ful quantities.

The lattice depth is linearly ramped up with a ramp speed $\alpha \equiv dV/d\tau$. In particular, $V(\tau) = V_0 + \alpha \tau$. As lattice depth is increased, the fraction $\frac{U}{t}$ increases exponentially in time (see eqs. (\ref{eq:TunellingVsLatticeDepth}) and (\ref{eq:InteractionVsLatticeDepth})). Since most of the dynamics occur very early in the lattice ramp, we can Taylor-expand $U/t$ as a function of time and only keep up to linear order terms (to make use of the Landau Zener solution later on): 
\begin{align}
        \frac{U}{t}(\tau) &= \frac{k_L a}{\sqrt{2}} \exp{\left(2\sqrt{\frac{V(\tau)}{E_r}} \right)}\\
        &\approx \frac{U_0}{t_0} + \left[ \frac{d}{d\tau}\left(\frac{U}{t}\right) \right]_{\tau = \tau_0} (\tau - \tau_0) \\ 
        &=\frac{U_0}{t_0} + \frac{U_0}{t_0} \frac{1}{\sqrt{E_r V(\tau_0)}}\left[\frac{d}{d\tau}V(\tau)\right]_{\tau=\tau_0} (\tau - \tau_0) \\
        &=\frac{U_0}{t_0} +  \frac{U_0}{t_0^2} \frac{\alpha}{\sqrt{E_r V(\tau_0)}} \tilde{\tau} \\
        &\equiv \frac{U_0}{t_0}+\alpha_{LZ}\tilde{\tau}.
\end{align}
Here we define a rescaled ramp speed $\alpha_{LZ}$ given by
\begin{equation}\label{alphaLZ}
    \alpha_{LZ}=\frac{U_0}{t_0^2}\frac{\alpha}{\sqrt{E_rV(\tau_0)}}
\end{equation}
To eliminate the constant offset in $\frac{U}{t}(\tau)$, we define $\tau^* = \tilde{\tau} + \tau_0^*$ where $\tau_0^* = \frac{1}{\alpha_{LZ}}\frac{U_0}{t_0}$ such that $U/t = \alpha_{LZ} \tau^*$. 
The 2nd order differential equation for $c_3$ can then be written as 
\begin{equation}
    \Ddot{a}_1 +i \alpha_{LZ} \tilde{\tau} \dot{a}_1+(g^2+i\alpha_{LZ} ) a_1=0,
\end{equation}
where $\dot{a}_1 = \frac{d a_1}{d\tau^*}$. We can eliminate the $\dot{a}_1$ term by going to the rotating frame, $a_1(\tau^*)=\Tilde{a_1}(\tau^*)e^{-i\frac{\alpha_{LZ}}{4}(\tau^*)^2}$. The differential equation becomes
\begin{equation}
    \Ddot{\tilde{a}}_1 + \left(g^2+\frac{i\alpha_{LZ}}{2} + \frac{\alpha_{LZ}^2}{4}(\tau^*)^2\right) \tilde{a}_1=0.
\end{equation}
From here we use the linear transformation $\tau^*\to z(\tau^*)=e^{i\frac{\pi}{4}}(\alpha_{LZ})^{1/2}\tau^*$ to obtain the standard form of the parabolic cylinder equations \cite[Ch.~12]{parabolic}
\begin{equation}\label{parabolic}
    \frac{d^2}{dz^2}\Tilde{a}_1(z)-\Tilde{a}_1(z)\left(\frac{1}{4}z^2+a\right)=0,
\end{equation}
with $a=\frac{ig^2}{\alpha_{LZ}}-\frac{1}{2}$.
There are two linear independent even and odd solutions to this differential equation given by the confluent hypergeometric functions $M$ \cite{AbStegun,parabolic}
\begin{equation} \label{eq:WeberSolutions}
    \begin{split}
        \text{even: } f_1(z)=e^{-\frac{1}{4}z^2} M\left(\frac{1}{2}a+\frac{1}{4},\frac{1}{2},\frac{1}{2}z^2\right)\\
        \text{odd: }f_2(z)=ze^{-\frac{1}{4}z^2}M\left(\frac{1}{2}a+\frac{3}{4},\frac{3}{2},\frac{1}{2}z^2\right)
    \end{split}
\end{equation}

Thus the coefficient $\Tilde{a}_1(z)=\tilde{A}_1 f_1(z)+\tilde{A}_2 f_2(z)$ is the superposition of these two solutions where $\tilde{A}_1$ and $\tilde{A}_2$ are given by the initial conditions of the problem. Going back to the original frame, we have $a_1(z) = \tilde{A}_1 f_1(z) e^{-\frac{z^2}{4}} + \tilde{A}_2 f_2(z)e^{-\frac{z^2}{4}}$. If $U_0/t_0$ is sufficiently large, $|z|$ is always large throughout the ramp. In this case, it is possible to write out an analytical expression for the asymptotic behaviour of the solutions eq. (\ref{eq:WeberSolutions}) for large $z$. This series expansion of the confluent hypergeometric functions for large $|z|$ is given by \cite[Ch.~13.7]{parabolic}:
\begin{equation}
    M\left(\tilde{a},\tilde{b},\tilde{z}\right)\sim\Gamma(\Tilde{b})\left(\frac{e^{i\pi \tilde{a}} \tilde{z}^{-\tilde{a}}}{\Gamma(\tilde{b}-\tilde{a})}+\frac{e^{\tilde{z}}\tilde{z}^{\tilde{a}-\tilde{b}}}{\Gamma(\tilde{a})}\right).
\end{equation}
We plug this asymptotic expansion into the even and odd solutions eq. (\ref{eq:WeberSolutions}), absorb all constants into new coefficients $B_1$ to $B_4$ and find:
\begin{align}
    f_1(z) &= e^{-\frac{1}{4}z^2} M\left(\frac{i g^2}{2\alpha_{LZ}},\frac{1}{2},\frac{z^2}{2} \right) \notag \\
    \Rightarrow f_1(z) &\approx B_1 e^{\left(-\frac{z^2}{4} -\frac{i g^2}{\alpha_{LZ}}\ln{|z|}\right)} + \frac{B_2}{z} e^{\left(\frac{z^2}{4} + \frac{i g^2}{\alpha_{LZ}}\ln{|z|}\right)} \\
    f_2(z) &= ze^{-\frac{z^2}{4}} M\left(\frac{i g^2}{2\alpha_{LZ}} + \frac{1}{2},\frac{3}{2},\frac{z^2}{2} \right) \notag \\
    \Rightarrow f_2(z) &\approx B_3 e^{\left(-\frac{z^2}{4} -\frac{i g^2}{\alpha_{LZ}}\ln{|z|}\right)} + \frac{B_4}{z} e^{\left(\frac{z^2}{4} + \frac{i g^2}{\alpha_{LZ}}\ln{|z|}\right)}
\end{align}
Writing out $a_1(\tau^*)$ using these asymptotic expansions, we get
\begin{equation}
    a_1(\tau^*) = A_1 e^{-\frac{i \alpha_{LZ}}{2} (\tau^*)^2} e^{-\frac{ig^2}{\alpha_{LZ}}\ln{\left(\sqrt{\alpha_{LZ}}\tau^*\right)}}
    + A_2\frac{1}{\tau^*} e^{\frac{ig^2}{\alpha_{LZ}}\ln{\left(\sqrt{\alpha_{LZ}}\tau^*\right)}} 
\end{equation}
where we defined new constants $A_1 = (\tilde{A}_1 B_1 + \tilde{A_2} B_3)$ and $A_2 = (\tilde{A}_1 B_2 + \tilde{A_2} B_4) \frac{e^{-i\pi/4}}{\sqrt{\alpha_{LZ}}}$. We can now obtain the full solution by solving for $A_1$ and $A_2$. The first constraint on $A_1$ and $A_2$ comes from the initial state being the ground state:
\begin{align} 
        a_1(\tau=0) &= a_1^0 \notag \\ 
        \Rightarrow a_1^0 &= A_1 e^{-\frac{i \alpha_{LZ}}{2} (\tau_0^*)^2} e^{-\frac{i g^2}{\alpha_{LZ}}\ln{\left(\sqrt{\alpha_{LZ}}\tau_0^*\right)}}
    + A_2\frac{1}{\tau_0^*} e^{\frac{i g^2}{\alpha_{LZ}}\ln{\left(\sqrt{\alpha_{LZ}}\tau_0^*\right)}} \notag\\
    \Rightarrow A_2\frac{1}{\tau_0^*} e^{\frac{i g^2}{\alpha_{LZ}}\ln{\left(\sqrt{\alpha_{LZ}}\tau_0^*\right)}} &= a_1^0 - A_1 e^{-\frac{i \alpha_{LZ}}{2} (\tau_0^*)^2} e^{-\frac{i g^2}{\alpha_{LZ}}\ln{\left(\sqrt{\alpha_{LZ}}\tau_0^*\right)}}
     \label{eq:InitialCondition1}
\end{align}
And the 2nd constraint on $A_1$ and $A_2$ is given by the equation of motion, eq. \ref{eq:TwoSiteEOM}:
\begin{alignat}{2}
    \left[ \frac{\partial a_1}{\partial \tau^*} \right]_{\tau_0^*}
    &= i g a_0^0 - i\frac{U_0}{t_0}a_1^0 \notag \\
    \Rightarrow \left(i g a_0^0 -i \frac{U_0}{t_0}a_1^0\right) 
    &= -A_1 e^{-\frac{i \alpha_{LZ}}{2} (\tau_0^*)^2} e^{-\frac{i g^2}{\alpha_{LZ}}\ln{\left(\sqrt{\alpha_{LZ}}\tau_0^*\right)}} \left( i \alpha_{LZ} \tau_0^* + \frac{ig^2}{\alpha_{LZ}} \frac{1}{\tau_0^*} \right) \notag\\
    & \quad + A_2\frac{1}{\tau_0^*} e^{\frac{i g^2}{\alpha_{LZ}}\ln{\left(\sqrt{\alpha_{LZ}}\tau_0^*\right)}} \left( \frac{i g^2}{\alpha_{LZ}}\frac{1}{\tau_0^*} -\frac{1}{\tau_0^*}\right) \label{eq:SolvingForA1Step2} \\
    \Rightarrow \left(i g a_0^0 -i \frac{U_0}{t_0}a_1^0\right) - a_1^0 \left( \frac{i g^2}{\alpha_{LZ}}\frac{1}{\tau_0^*} - \frac{1}{\tau_0^*} \right)
    &= - A_1 e^{-\frac{i \alpha_{LZ}}{2} (\tau_0^*)^2} e^{-\frac{i g^2}{\alpha_{LZ}}\ln{\left(\sqrt{\alpha_{LZ}}\tau_0^*\right)}} \left( i \alpha_{LZ} \tau_0^* 
     + \frac{2 i g^2}{\alpha_{LZ}} \frac{1}{\tau_0^*} - \frac{1}{\tau_0^*} \right) \notag \\
    \Rightarrow A_1 e^{-\frac{i \alpha_{LZ}}{2} (\tau_0^*)^2} e^{-\frac{i g^2}{\alpha_{LZ}}\ln{\left(\sqrt{\alpha_{LZ}}\tau_0^*\right)}} 
    &= \frac{\left(ig a_0^0 -i \frac{U_0}{t_0}a_1^0\right) - a_1^0 \left( \frac{i g^2}{\alpha_{LZ}}\frac{1}{\tau_0^*} - \frac{1}{\tau_0^*} \right)}
    {\left( i \alpha_{LZ} \tau_0^*  + \frac{2 i g^2}{\alpha_{LZ}} \frac{1}{\tau_0^*} - \frac{1}{\tau_0^*} \right)} \label{eq:InitialCondition2} \\
    \Rightarrow  A_2\frac{1}{\tau_0^*} e^{\frac{i g^2}{\alpha_{LZ}}\ln{\left(\sqrt{\alpha_{LZ}}\tau_0^*\right)}} 
    &= \frac{ig a_0^0  +\frac{i g^2}{\alpha_{LZ}}\frac{1}{\tau_0^*}  a_1^0}
    {\left( i \alpha_{LZ} \tau_0^*  + \frac{2 i g^2}{\alpha_{LZ}} \frac{1}{\tau_0^*} - \frac{1}{\tau_0^*} \right)}
\end{alignat}
In the second step eq. (\ref{eq:SolvingForA1Step2}) we plugged in the relation given by eq. (\ref{eq:InitialCondition1}). This fully specifies the solution for all times. 

For our purposes, we are interested in the density of doublons in the system defined as
$$ \langle \hat{\rho}_D \rangle = \frac{1}{N} \sum_{i}{\bra{\Psi} \hat{n}_{i\uparrow} \hat{n}_{i\downarrow} \ket{\Psi} }, $$
where $N$ is the number of sites and index $i$ runs over all lattice sites. For the two-site system, $N=2$ and $i\in \{L,R\}$. 
Conveniently in our chosen basis, $\bra{\phi_0} \hat{\rho}_D \ket{\phi_0} = 0$ and $\bra{\phi_1} \hat{\rho}_D \ket{\phi_1} = 1/2$. 
Thus the doublon density is simply given by $\langle \rho_D \rangle = |c_1|^2/2$, i.e. half the population of the state $\ket{\phi_1}$. Knowing the full solution to $a_1(\tau)$, we can easily compute the doublon density as a function of time. In particular, at the end of the ramp, $\tau^*\to \infty$. Hence, the doublon density at the end of the ramp is given by
\begin{equation}
    \langle\hat{\rho}_D\rangle = \frac{1}{2}|a_1(\tau^*\to \infty)|^2 = \frac{1}{2}|A_1|^2 
\end{equation}
since the term with $A_2$ falls off as $1/\tau^*$ and vanishes for large $\tau^*$. Simplifying the expression in eq. (\ref{eq:InitialCondition2}), we find:
\begin{equation}\label{eq:A1}
    |A_1|^2 = \frac{ \left( a_1^0 \alpha_{LZ} \frac{t_0^2}{U_0^2}\right)^2 + a_0^2\left( 1 + g^2\frac{t_0^2}{U_0^2}\right)^2 \left( \frac{a_1^0}{a_0^0}  - \frac{g\frac{t_0}{U_0}}{ 1 + g^2\frac{t_0^2}{U_0^2}}\right)^2}
    {\left( \alpha_{LZ} \frac{t_0^2}{U_0^2}\right)^2 + \left(1 + 2g^2\frac{t_0^2}{U_0^2} \right)^2}.
\end{equation}
We wish to show that the 2nd term in the numerator of eq. (\ref{eq:A1}) can be neglected when $U_0 \gg t_0$. To see this, we first identify $a_1^0/a_0^0 = \tan{\theta}$ where $\theta$ is given by eq. (\ref{eq:TwoSiteEigenstates}). We can then Taylor expand the 2nd term in orders of $t_0/U_0$ when $t_0 \ll U_0$: 
\begin{align}
    \left( \frac{a_1^0}{a_0^0}  - \frac{g\frac{t_0}{U_0}}{ 1 + g^2\frac{t_0^2}{U_0^2}}\right)^2 
    &= \left( \frac{(-1 + \sqrt{1 + 16 \frac{t_0^2}{U_0^2}}}{\frac{4t_0}{U_0}} - \frac{\frac{2t_0}{U_0}}{ 1 + 4\frac{t_0^2}{U_0^2}}\right)^2 
    \approx \left(-32 \frac{t_0^5}{U_0^5} + O\left(\frac{t_0^6}{U_0^6}\right)\right)^2
\end{align}
Thus the 2nd term in the numerator of eq. (\ref{eq:A1}) is of the order $O((t_0/U_0)^{10})$ and can thus be neglected when $U_0 \gg t_0$. Plugging in the value of $\alpha_{LZ}$ in terms of the ramp speed $\alpha$, we find
\begin{align}
    |A_1|^2 \approx (a_1^0)^2 \frac{ \alpha^2}
    {\alpha^2 + U_0^2 E_r V_0 \left(1 + \frac{8t_0^2}{U_0^2} \right)^2}.
\end{align}
We define a critical ramp speed $\alpha^*$ where $|A_1|^2$ has an inflection point:
\begin{equation}
    \alpha^* = U_0 \sqrt{E_r V_0} \left(1 + \frac{8t_0^2}{U_0^2} \right) 
\end{equation}
This gives us the doublon density $\langle\hat{\rho}_D\rangle$ as a function of ramp speed $\alpha$ as:
\begin{align}
    \langle \hat{\rho}_D \rangle = \frac{(a_1^0)^2}{2} \frac{ \alpha^2}
    {\alpha^2 + (\alpha^*)^2}.
\end{align}

\newpage
\section{Numerical simulations} \label{app:Numerics}
The numerical simulations in this work were performed using code developed in house in the Python programming language. The numerical libraries Numpy and Scipy were heavily used for efficient computations, as well as python multiprocessing to parallelize computations across CPU cores. In addition, we used Cython to precompile python functions into C-code to greatly speed up key functions that required branching (loops, if statements). With this code, we are able to study systems of up to 14 sites in the Fermi-Hubbard model on a standard desktop computer.

We work in the fixed particle number and fixed magnetization sector, i.e. fixed $N_{\uparrow}$ and $N_{\downarrow}$. The Hilbert space dimension of the largest magnetization sector scales as $\sim 2^{2N}/N$. In the fixed magnetization sectors, all operators in the Hilbert space can be constructed from the generalized ``hopping" operators $\hat{c}^{\dagger}_{i\sigma} \hat{c}_{j\sigma}$ which conserve spin and particle number. All terms in the Hamiltonian as well as all observables can be represented as sparse matrices using a matrix representation of these hopping operators. We choose the Fock basis of the Fermi-Hubbard model as the basis for writing out explicit matrix representations of operators in our numerical simulations. Since we are working with fermionic particles, we need to be careful about the order of the creation operators $\hat{c}^{\dagger}_{i\sigma}$ used to define the Fock basis states. To be consistent, we use the rule that all spin-up operators lie to the left of spin-down operators and the site indices are arranged in ascending order. 
Once we compute matrix representations of the hopping operators, we no longer have to worry about fermionic signs, since the hopping operators $\hat{c}^{\dagger}_{i\sigma} \hat{c}_{j\sigma}$ behave as bosons (they contain an even number of fermions).

\subsection{Two site numerics}
For Fig. \ref{fig:Doublons}(b), we simulate the two-site Fermi-Hubbard system at half-filling using exact-diagonalization (ED) and time-evolution with the time-dependent Fermi-Hubbard Hamiltonian (eq. (\ref{eq:Hubbard2})). 
Throughout this work we use an initial lattice depth of $V_0^{\text{initial}} = 8 E_r$. 
We calculate tunnelling strength $t$ and interaction $U$ from the lattice depth $V_0$ using eqs. (\ref{eq:TunellingVsLatticeDepth}) and (\ref{eq:InteractionVsLatticeDepth}). 
We tune the initial ratio $U_0/t_0$ by varying the $s$-wave scattering length $a$ from eq. (\ref{eq:InteractionVsLatticeDepth}). We can treat $a$ as a free parameter because in experiment, we can vary the scattering length using a Feshbach resonance. 
We work with two different values of the initial interaction to tunnelling ratio $U_0/t_0 = 8$ and $U_0/t_0 = 16$. 
We find the ground state of the Fermi-Hubbard Hamiltonian using ED and use that as the initial state.
We ramp up the lattice depth linearly in time at a rate $\alpha$ up to a final lattice depth $V_0^{\rm{final}} = 48 E_r$ again to match the conditions of \cite{Mazurenko2017}. 
We calculate the doublon-density in the final state by directly computing the expectation value of the doublon operator in the final state.

\subsection{Ground state numerics}
For Figs. \ref{fig:Doublons}(c), \ref{fig:SpinCorr}(a) and \ref{fig:LatticeRampUnitary}(a), we work with system size of 12 sites in a 1D periodic chain or 2D cluster ($4\times3$, periodic along $x$ and open boundary conditions along $y$). We use the ground state of the Fermi-Hubbard Hamiltonian with $U_0/t_0=8$ as the initial state. The ground state is numerically computed using a built-in sparse diagonalization algorithm (scipy.linalg.sparse.eigsh) based on the Lanczos method \cite{Lanczos1950}. 
To perform the time-evolution with the time-dependent Fermi-Hubbard Hamiltonian, we trotterize the time-evolution operator as 
\begin{equation}
\begin{split}
\hat{U}_{\text{ramp}} &= T\left[\exp{\left(-i \int_0^\tau{ d\tau' \hat{H}_{\rm{FH}}(\tau')}\right)}\right] \\
&\approx \prod_{\substack{n=0,\\ \tau' = n \frac{\tau}{\Delta\tau}}}^{\tau/\Delta \tau}{\exp{\left(-i  \Delta\tau \hat{H}_{\rm{FH}}(\tau')\right)}}\\
&\approx \prod_{\substack{n=0,\\ \tau' = n \frac{\tau}{\Delta\tau}}}^{\tau/\Delta \tau}{\exp{\left(-\frac{i \Delta\tau}{2} \hat{H}_{\rm{int}}(\tau')\right)} \exp{\left(-i \Delta\tau \hat{H}_{\rm{kin}}(\tau')\right)} \exp{\left(-\frac{i \Delta\tau}{2} \hat{H}_{\rm{int}}(\tau')\right)}}
\end{split}
\end{equation}
where $\hat{H}_{\rm{kin}}$ and $\hat{H}_{\rm{int}}$ represent the kintetic energy and interaction term of the Fermi-Hubbard Hamiltonian given by the first and second term of eq. (\ref{eq:FermiHubbardHamiltonian}) respectively. Since we perform the numerical simulations in the Fock basis of the Fermi-Hubbard Hamiltonian, $\hat{H}_{\rm{int}}$ is a diagonal operator and thus so is $\exp{\left(-\frac{i \Delta\tau}{2} \hat{H}_{\rm{int}}(\tau')\right)}$. On the other hand, $\hat{H}_{\rm{kin}}$ is off-diagonal in the Fock basis. Instead of computing the full matrix corresponding to $\exp{\left(-i \Delta\tau \hat{H}_{\rm{kin}}(\tau')\right)}$, we directly compute the action of the operator on the wavefunction using built-in sparse matrix functions (scipy.sparse.linalg.expm\_multiply).
We choose a step size $\Delta \tau$ small enough that the trotter error is negligible which is confirmed by checking convergence of the wavefunction as a function of decreasing step size. 

We also make comparisons of the time-evolved Fermi-Hubbard state with the ground state of the $t-J-3s$ model. The ground state of the $t-J-3s$ model (see eq. (\ref{eq:tJmodelHamiltonian})) is numerically computed using sparse diagonalization by writing the $t-J-3s$ Hamiltonian in the Fermi-Hubbard Fock basis (i.e. replacing $\tilde{c}_{i\sigma}$ with $c_{i\sigma}$). We use the Fermi-Hubbard basis to write the $t-J-3s$ Hamiltonian since the goal of the lattice ramp protocol is to perform the Schrieffer-Wolff basis rotation which maps $\hat{\tilde{c}}_{i\sigma} \to \hat{c}_{i\sigma}$. 

\subsection{Finite temperature numerics}
In Fig. \ref{fig:SpinCorr}(b), we perform numerical simulations at finite temperature $T\ll U_0$ in a system of $3\times3$ sites at half filling and with open boundary conditions. For the finite temperature computations, the initial state of the system would be a thermal ensemble of the Fermi-Hubbard model with eignestate populations given by the Boltzmann distribution. Furthermore, we need to consider eigenstates of the Hamiltonian in all the magnetization sectors, not just the largest sector as is the case for the ground state. 

We perform wavefunction time-evolution as opposed to density matrix time-evolution because of computer memory constraints. For each magnetization sector, we use ED to find all the eigenstates in that sector. We can use full ED since the Hilbert space is small enough for a $3\times3$ system. 
Starting with a given eigenstate $\ket{\Psi_n}$ with energy $E_n$, we perform the numerical time-evolution same as described above and compute observables $\langle \hat{O}\rangle_n$ in this time-evolved state. We then assign the observable $\langle \hat{O}\rangle_n$ a coefficient $P_n$ given by the Boltzmann weight of the eigenstate $\ket{\Psi_n}$, i.e. $P_n = e^{-E_n T} / Z(T)$ where $T$ is the temperature of the initial state and $Z(T)$ is the partition function for that temperature. We sum up the contributions from the lowest $\sim 150$ eigenstates from each magnetization sector to compute the observable in the finite temperature time-evolved state. We find that $\sim 150$ states were enough for the numerical results to converge. 

For Fig. \ref{fig:SpinCorr}(b), we plot the full $\langle \hat{\vec{S}}_i\cdot \hat{\vec{S}}_j \rangle$ spin correlator instead of just the $z$-component $\langle S^z_i S^z_j \rangle$ because for a fixed magnetization sector, the $z$-component correlator is non-monotonic with respect to temperature. While the Fermi-Hubbard Hamiltonian has SU(2) total spin rotation symmetry, by choosing to work in a fixed magnetization sector, we break the SU(2) symmetry and only make use of the U(1) symmetry arising from charge conservation. For ground state numerics, it is still sufficient to only look at $\langle S^z_i S^z_j \rangle$ correlations since in that case, $\langle \hat{S}^z_i \hat{S}^z_j \rangle = \langle \hat{S}^x_i \hat{S}^x_j \rangle = \langle \hat{S}^y_i \hat{S}^y_j \rangle = \langle \hat{\vec{S}}_i\cdot \hat{\vec{S}}_j \rangle / 3$.

\end{document}